\documentclass[aps,prab,twocolumn,superscriptaddress]{revtex4-2}

\usepackage{graphicx}
\usepackage{dcolumn}
\usepackage{bm}
\usepackage{hyperref}
\usepackage{amsmath}
\usepackage{upgreek}

\begin{document}

\title{Longitudinal phase space manipulation using double emittance exchange to generate multi-color X-ray}

\author{Jimin Seok}
\email{linker123@unist.ac.kr}
\affiliation{UNIST, Ulsan, 44919, South Korea}
\affiliation{Argonne National Laboratory, Lemont, Illinois 60439, USA}
\author{Gwanghui Ha}
\affiliation{Argonne National Laboratory, Lemont, Illinois 60439, USA}
\author{John Power}
\affiliation{Argonne National Laboratory, Lemont, Illinois 60439, USA}
\author{Moses Chung}
\affiliation{UNIST, Ulsan, 44919, South Korea}

\date{\today}


\begin{abstract}
Generating temporally separated two X-ray pulses or even two pulses with different colors has been pursued for various X-ray experiments.  Recently, this concept is extended to generate multi-color X-ray pulses, and a few approaches have been proposed. We introduce one of possible new ways to generate multi-color X-ray using a longitudinal phase space (LPS) modulator and a manipulator. In this example, a wakefield structure and double-emittance exchange beamline are used as the LPS modulator and the LPS manipulator, respectively. In this  way, we can generate multiple bunches having designed energy and time separations. These separations can be adjusted for each application differently. This paper describes the principle of the method and its feasibility.
\end{abstract}

\maketitle


\section{Introduction}

Coherent X-ray source is one of the most important equipment for modern science. Its application covers various fields such as material science and biology \cite{gisriel2019membrane,xu2014single,alonso2020femtosecond,hirata2014determination,dean2016ultrafast}. Since the first hard X-ray free electron laser facility started its operation \cite{mcneil2009first}, extensive researches to provide better X-ray conditions for applications have been carried out. One of the research directions was generating two X-ray pulses having time separation or even energy separation, so called two-color X-ray \cite{PhysRevLett.84.2861,PhysRevLett.110.134801,PhysRevLett.113.254801,PhysRevAccelBeams.23.030703,hara2013two,marinelli2015high,PhysRevResearch.2.042018}. Recently these researches were extended to generate multi-color X-rays. So far, only a few different methods have been proposed \cite{lutman2016fresh,hemsing2019soft,qiang2011generation,xiang2012mode}, and further development is necessary to  explore possible performance improvements.

We introduce a new method to generate multi-color X-ray in this paper which uses beam's longitudinal phase space manipulation. This method consists of two key elements, longitudinal phase space (LPS) modulator and LPS manipulator. The modulator applies sinusoidal modulation on a single beam's LPS. Then, the manipulator controls this modulation to generate spectral and temporal bunching. Any modulation method can be used for this purpose, but here we assume to use wakefield structure. Also, we use double emittance exchange (EEX) beamline as the LPS manipulator.

The LPS modulator should provide a modulation of which wavelength is shorter than the bunch length. Here the number of periods within the bunch determines the number of bunches from temporal/spectral bunching. The modulation amplitude is one of the factors determining bunching level. In the case of wakefield structures, the modulation wavelength is determined by geometry and material \cite{PhysRevLett.61.2756,PhysRevSTAB.3.101302,BANE2012106}, and tunable frequency control is available for dielectric slab structures \cite{PhysRevE.56.4647,PhysRevLett.108.244801}. The modulation amplitude can be controlled by both the design of the structure and the charge level of the beam passing through the structure. If one uses collinear wakefield acceleration concept \cite{bane1985col,PhysRevSTAB.15.011301,PhysRevLett.120.114801}, the amplitude can be adjusted by controlling the charge level of a wakefield-driving beam.

In the case of the LPS manipulator, double EEX \cite{Ha:FEL2017-TUP054,zholents2011new} is one of possible options. The double EEX beamline enables the control of LPS via transverse manipulations. As shown in many references \cite{PhysRevAccelBeams.21.014401,seok2019suppression,seok2018sub,PhysRevSTAB.14.022801,Ha:IPAC2019-TUPGW089,PhysRevLett.105.234801,PhysRevAccelBeams.19.121301}, longitudinal properties can be manipulated using quadrupole magnets. This provides variable  ${R}_{55}$, ${R}_{56}$, ${R}_{65}$ and ${R}_{66}$,  which enables bunching of particles in both time and energy. It controls generated bunch's separations simultaneously.

The following sections provide the principle of the proposed method and experimental data showing its feasibility. Note that this experimental data is what we observed during other experiment. Thus, it does not show all aspects of this new method.

\section{Principle of new method}

\begin{figure*}[t]
\centerline{\includegraphics[width=0.5\textwidth,keepaspectratio=true]{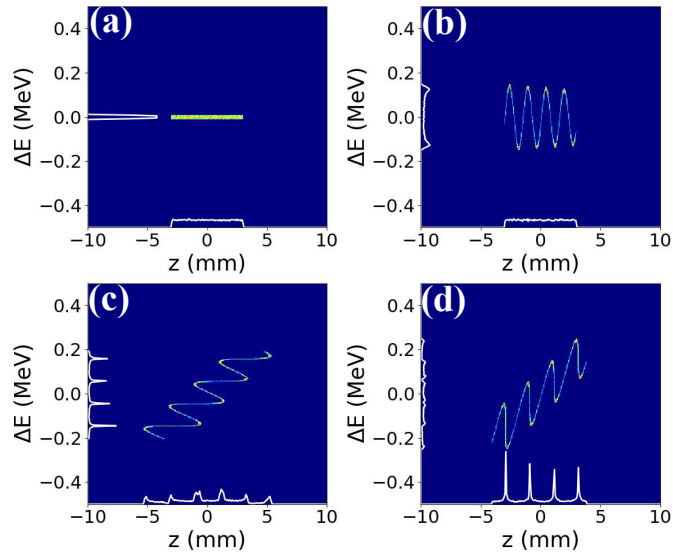}}
\caption{\textbf{Longitudinal phase spaces of numerically tracked beam.} An artificially generated beam travels through a 200 GHz dielectric structure and double EEX beamline to illustrate density and spectral bunching cases. Panel (a) shows the initial beam's LPS with uniform density distribution. Its full width is 6 mm to fit in four modulation periods. Panel (b) shows the LPS after the dielectric structure. Panels (c) and (d) correspond to spectral bunching and density bunching results, respectively. Particle tracking through a double EEX beamline is performed by using transfer matrix.}
\label{fig:stream}
\end{figure*}

In the case of conventional methods, they first introduce a sinusoidal modulation to beam's LPS. Then, proper $R_{56}$ from anisochromatic beamline such as chicane \cite{antipov-2013-a} converts this modulation to density modulation. Due to the energy difference that energy modulation introduced, particles having higher energy catch up preceding low energy particles via $R_{56}>0$. They move to zero-crossing points of the sine curve, and the density profile can be bunched when an appropriate $R_{56}$ is introduced. The new method we introduce uses the same principle but we use all $R_{55}$, $R_{56}$, $R_{65}$, and $R_{66}$ to generate either temporal or spectral bunching, and control their separation via EEX beamline. Equations given later in this section confirm this process and several simple but rough estimation of its characteristics. We assume a beam having zero longitudinal emittance with a linear chirp. Also, only its itnitial LPS was used to describe final LPS to simplify the discussion.

The first step of the method is applying sinusoidal modulation to the beam. When the beam passes through a high impedance medium such as dielectric tube, it generates wakefield. When the wakefield modulation frequency is high enough, this wakefield can introduce sinusoidal energy modulation to the beam. It can be written as,
\begin{equation} \label{eq_modulation}
\delta(z) = \frac{G_{m}L_{m}}{ E_0} \sin\left(\frac{\omega z}{c}\right) + Hz,
\end{equation}
where $G_m$ is the averaged gradient of the modulator, $L_m$ is the length of the modulator, $\omega$ is the angular frequency of the modulator's wakefield, $E_0$ is the reference beam energy, and $H$ is the initial longitudinal chirp that we call as macro-chirp. Because we know that the bunching happens near the zero-crossings, it is reasonable to observe only a half wavelength area nearby the zero-crossings (i.e. $-\frac{\lambda}{4} < z-n\lambda < +\frac{\lambda}{4}$) where the bunching happens. Then, Eq.~\eqref{eq_modulation} can be linearly approximated as,
\begin{equation}
\delta_n \simeq h\bar{z} + H(\bar{z}+n\lambda),
\end{equation}
where $|\bar{z}|<\frac{\lambda}{4}$ and $h \equiv \frac{\omega G_mL_m}{cE_0}$. We will call this $h$ as micro-chirp.

Next, this beam enters a double EEX beamline which has certain ${R}_{55}$, ${R}_{56}$, ${R}_{65}$ and ${R}_{66}$. Note that there will be quadrupole magnets in the middle of the double EEX beamline, and they can control these transfer matrix elements. Thus, the final LPS coordinates after the double EEX beamline become,
\begin{eqnarray}
z_{n,f} = R_{55}(\bar{z}+n\lambda) + R_{56}h\bar{z} + R_{56}H(\bar{z}+n\lambda), \\
\delta_{n,f} = R_{65}(\bar{z}+n\lambda) + R_{66}h\bar{z} + R_{66}H(z+n\lambda).
\end{eqnarray}
These simple equations tell us that longitudinal density can be bunched when the energy gain from modulator ($G_mL_m$) is equal to $-\frac{R_{55}+R_{56}H}{ R_{56}}\frac{cE_0}{\omega}$. Typically, the macro-chirp is much smaller than the micro-chirp, so it can be further simplified to $-\frac{R_{55}}{ R_{56}}\frac{cE_0}{\omega}$. Here, the bunch-to-bunch separation can be found from terms having no $\bar{z}$ dependency [i.e. $\lambda(R_{55}+R_{56}H)$]. Because both the bunching condition and the separation are determined by $R_{55}$ and $R_{56}$, quadrupole combinations are able to bunch the beam with an arbitrary separation. The same argument holds for the spectral bunching. Here, the bunching condition will be $-\frac{R_{65}+R_{66}H}{ R_{66}}\frac{cE_0}{\omega}$ and the energy deviation will be $\lambda (R_{65}+R_{66}H)E_0$.

This concept can be easily confirmed by tracking particles using transfer matrix (see FIG.~\ref{fig:stream}). For a numerical example, we assumed a single mode dielectric structure whose frequency is 200 GHz. Quadrupole magnets were assumed to be located in the middle of double EEX beamline to control ${R}_{55}$, ${R}_{56}$, ${R}_{65}$ and ${R}_{66}$. Their strengths were optimized to produce density and spectral bunchings. To apply the wakefield more realistically, the wake function of the structure was convoluted with beam's longitudinal density distribution. Two particle distributions having different charge were generated to represent wakefield-drive beam and target beam, and their charges were 2 nC and 200 pC, respectively. The beam energy of 50 MeV was assumed. The target beam had uniform longitudinal density distribution and its length was 4 times of the modulation wavelength (i.e., 6 mm); see FIG.~\ref{fig:stream}. 

After the target bunch passes through the structure, it achieved a sinusoidal energy modulation as shown in the panel (b). This modulation is successfully converted to spectral and density bunchings as shown in panels (c) and (d).  The macro-chirp was zero, and the micro-chirp from the modulation was -10.98 m$^{-1}$. Quadrupole magnets were optimized to generate bunchings, and the double EEX beamline provided \textit{R}-terms as given in Table \ref{tab}. Because the macro-chirp is zero, the bunching condition was 11.08 m$^{-1}$ for both cases, which shows good agreement with the micro-chirp calculated from the distribution in FIG.~\ref{fig:stream}(b). Also, we expect that the density and the energy separations would be 2 mm and 100 keV, respectively. These also show good agreements with FIGS.~\ref{fig:stream}(c) and (d).

\begin{table}[]
    \caption{Matrix element used for numerical tracking. $R_{55}$ and $R_{56}$ are matrix elements for density bunching, and the other two elements for the density bunching case are not listed in this table. Similarly, $R_{65}$ and $R_{66}$ in this table are matrix elements the spectral bunching, and the other two elements are not listed.}
    \begin{ruledtabular}
    \begin{tabular}{c c c c}
    $R_{55}$ & $R_{56}$ & $R_{65}$ & $R_{66}$  \\
    \hline
    1.33 & 0.12 & 1.33 & 0.12
    \end{tabular}
    \label{tab}
\end{ruledtabular}
\end{table}

\section{Experimental data showing feasibility of the method}

\begin{figure}[htb]
\centerline{\includegraphics[width=0.5\textwidth,keepaspectratio=true]{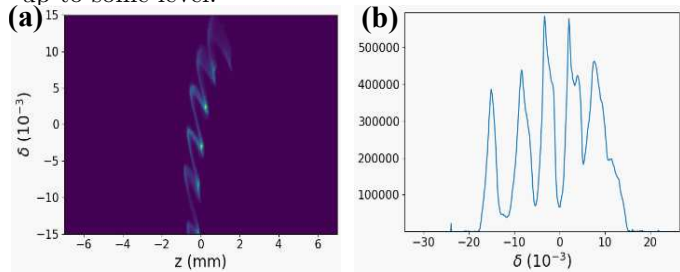}}
\caption{\textbf{Longitudinal phase space and corresponding spectral profile from transverse wiggler experiment.} Panel (a) shows measured longitudinal phase space after a double EEX beamline. Panel (b) shows spectral profile of the beam corresponding the panel (a).}
\label{fig:stream2}
\end{figure}

Argonne wakefield accelerator facility recently performed an experiment to demonstrate the beam manipulation concept using a transverse wiggler (details can be found from \cite{Ha:IPAC2019-TUPGW089}). Although this experiment implies sinusoidal modulation to the transverse phase space and uses a single EEX beamline, its underlying principle is the same as the method introduced in this paper. A sinusoidal modulation is introduced into the phase space, and \textit{R}-terms of the EEX beamline are controlled by quadrupole magnets to bunch the beam and control its separation. While the main goal of this experiment is to demonstrate the concept of the transverse wiggler, they observed one interesting result shown in FIG.~\ref{fig:stream2} at certain quadrupole setting. Both LPS image and corresponding spectral profile show clear bunching in the energy axis. The beam was originally generated from a photocathode gun, and it had a smooth energy spectrum. However, the modulation introduced by the modulator and proper setting of the following EEX beamline provided spectral bunching as we expected from the previous section.  This is a clear evidence of the feasibility of the new method.

Note that the fractional energy separation in FIG.~\ref{fig:stream2} is about 1\%, and it corresponds to 400 keV. Each micro-bunch is not critically compressed. The energy spread of each micro-bunch can be further reduced by adjusting quadrupole setup. While the bunch is compressed in density in this data, the bunch length can also be controlled up to some level.

\section{Conclusions}
We introduced a new method to generate multi-energy beam that can be used to generate multi-color X-ray. This paper discussed the principle of the method, and one experimental data showing the feasibility of the method is provided. The strength of this method is its capability of bunching the single beam in both time and energy axes. Also, an arbitrary control of the separations is available.  The details of the method are not covered in this paper, but we can see that the separation of time and energy can be controlled simultaneously up to some level. We believe that this method can be one way to provide interesting beam configurations to light source users.

\bibliography{reference}

\end{document}